# Detecting Propagators of Disinformation on Twitter Using Quantitative Discursive Analysis

Mark M. Bailey, PhD[1]

[1]*Cyber Intelligence and Data Science, Oettinger School of Science and Technology Intelligence, National Intelligence University, Bethesda, MD*

## Abstract

Efforts by foreign actors to influence public opinion have gained considerable attention because of their potential to impact democratic elections.  Thus, the ability to identify and counter sources of disinformation is increasingly becoming a top priority for government entities in order to protect the integrity of democratic processes.  This study presents a method of identifying Russian disinformation bots on Twitter using centering resonance analysis and Clauset-Newman-Moore community detection.  The data reflect a significant degree of discursive dissimilarity between known Russian disinformation bots and a control set of Twitter users during the timeframe of the 2016 U.S. Presidential Election.  The data also demonstrate statistically significant classification capabilities (MCC = 0.9070) based on community clustering.  The prediction algorithm is very effective at identifying true positives (bots), but is not able to resolve true negatives (non-bots) because of the lack of discursive similarity between control users.  This leads to a highly sensitive means of identifying propagators of disinformation with a high degree of discursive similarity on Twitter, with implications for limiting the spread of disinformation that could impact democratic processes.

## Background

Efforts by foreign actors to influence public opinion via the spread of disinformation have been thrust into the spotlight recently because of their potential to disrupt democratic processes [1].  While the injection of propaganda and misleading information into public discourse by adversaries is not a new phenomenon, the magnitude of its effects and reach can be greatly enhanced by the connectedness of social media [2].  Thus, the ability to identify and counter sources of disinformation is increasingly becoming a top priority for government entities in order to protect election integrity.

Disinformation source detection and countermeasure development rely heavily on mathematical representations of social interactions and textual information. There is a significant amount of research in the area of social network analysis [3], and quantitative analysis of text [4], for user classification and event detection. Additionally, several methods have already been developed that can identify disinformation vectors, i.e., "bots," using artificial intelligence [5], [6]. One method of quantitative textual analysis – centering resonance analysis – was developed by Corman, *et al.*, and is a form of network text analysis that relies on elements of graph theory to build mathematical representations of object associations within bodies of text [7].

Building on work by Corman, *et al.* [7], this study applies centering resonance analysis to a body of aggregated tweets from a previously-identified set of Russian bots [8], as well as a set of randomly selected tweets as a control set [9]. By representing aggregated tweet text for each Twitter user as graphs of noun phrases, one can represent discursive similarity ("resonance") as the dot product of the centralities of common vertices, where each vertex represents a noun from a noun phrase (discursive object), and edges represent connections between nouns (object associations). By analyzing a connected graph of all users, where edges are defined by resonance between vertices (i.e., discursive similarity between users), the Clauset-Newman-Moore hierarchical agglomeration algorithm can be applied to identify discursive communities [10].

Because they exist for a singular purpose, Russian bots are likely to be very limited in the topics they discuss on Twitter relative to the general population of Twitter users. Thus, it is hypothesized that they will show greater discursive similarity with each other than with the general Twitter population. This study will demonstrate this effect and will show that Russian bots aggregate within distinct graph communities. Identified communities can be used to develop recursive algorithms for disinformation propagator bot detection.

## Methods

### Software

Python version 2.7/3.1 was used in this analysis. The NetworkX, Natural Language Tool Kit, and TextBlob Python libraries were used for network analysis and natural language processing, respectively.

### Approach

The method employed in this study leverages centering resonance analysis to quantify discursive similarity between Twitter users – either known Russian bots, or randomly selected control users [7]. A set of tweets from known Russian bots that were active during the 2016 Presidential election, as well as a body of control tweets from random users, were acquired [8], [9]. Tweet text was aggregated by user, and all concatenated text aggregates were preprocessed. Preprocessing included punctuation removal, lemmatization, and case normalization. After preprocessing, noun phrases were extracted from each corpus, and noun edges and vertices were enumerated. Graphical representations of noun phrases were then generated for each user as a discursive graph – a mathematical representation of the user's discursive object associations.

Graph resonances – a measure of user interaction – were then calculated for each user pair, and an association matrix was constructed. A graphical representation of the overall network, where edges between users represent discursive similarity – was then constructed and optimized for community detection. The entire process is outlined in Figure 1:

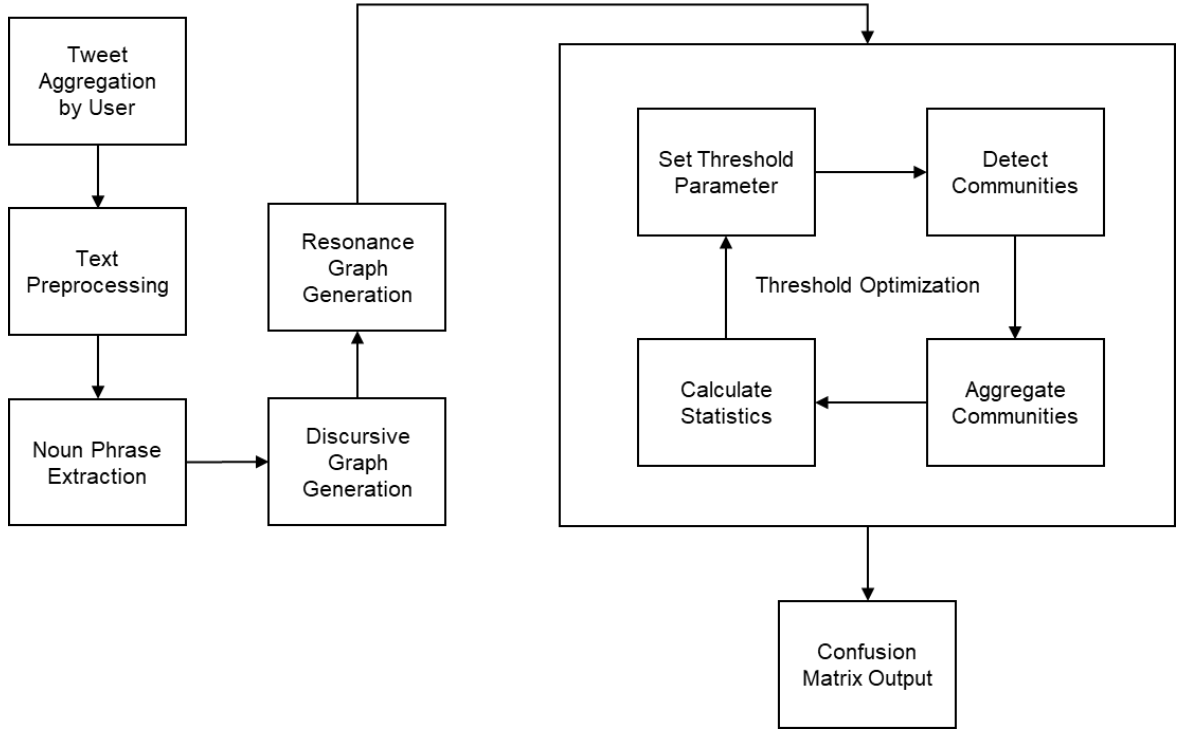

Figure 1: Process diagram.

In any graph, the betweenness centrality of a vertex (*v*) is defined as the sum of the fraction of all-pairs shortest paths that pass through the vertex, and is given in Equation 1:

$$I(v) = \sum_{s,t \in V} \frac{\sigma(s,t|v)}{\sigma(s,t)} \qquad (1)$$

In this equation, *V* is the set of vertices, *σ(s,t)* is the number of shortest *(s,t)* paths, and *σ(s,t|v)* is the number of those paths passing through some vertex *v* other than *s,t* [11]. Betweenness centralities were calculated for each user graph.

Word resonance – a measure of the discursive similarity between bodies of text – was calculated as the dot product of the betweenness centrality vectors of the common set of vertices between two user graphs (*A* and *B*), as follows [7]:

$$WR_{AB} = (I_A \cdot I_B) \ \forall \in \{V_A \cap V_B\} \qquad (2)$$

To construct a standardized measure, the resonance is normalized:

$$\overline{WR_{AB}} = \frac{WR_{AB}}{\sqrt{\sum_i (I_i^A)^2 \cdot \sum_j (I_j^B)^2}} \qquad (3)$$

Resonances were calculated for each user graph pair, and a matrix of resonances – a measure of interaction – was constructed as follows by calculating matrix element $m_{ij}$ for all user graph pairs:

$$m_{ij} = \begin{cases} \overline{WR_{ij}} \text{ if } i \neq j \\ 0 \text{ if } i = j \end{cases}, where\ m_{ij} \in [M] \qquad (4)$$

Graphs can be represented as an association matrix, where $a_{ij} = 1$ if and only if an edge exists between vertices *i* and *j*, otherwise $a_{ij} = 0$. By applying a threshold value, $\tau$, to the resultant matrix of resonance values, *[M]*, an association matrix, *[A]*, can be generated by calculating matrix element $a_{ij}$ for all $m_{ij}$:

$$a_{ij} = \begin{cases} 1 \text{ if } m_{ij} \geq \tau \\ 0 \text{ if } m_{ij} < \tau \end{cases}, where\ a_{ij} \in [A] \qquad (5)$$

Association matrices, as described in Equation 5, were generated over a range of threshold values to facilitate optimization. Once the discursive network was generated between users from an association matrix, clustering analysis using the Clauset-Newman-Moore hierarchical agglomeration algorithm was applied to identify discursive communities [10]. The Clauset-Newman-Moore algorithm uses network *modularity* – a measure comparing the edges within a community to the edges between communities – to determine the optimal number of communities within the graph. Because this study aims to develop a binary classification algorithm for Twitter users (*Russian Bot* or *Control User*), the resultant communities were pooled based on their ratios of bots to control users, such that communities containing a majority of bot users pooled together, and vice versa. Many of the communities detected using the Clauset-Newman-Moore algorithm were singular (isolated vertices), so only the communities of size > 1 were pooled using this method. Confusion matrices, and associated statistical measures, were then calculated for each threshold iteration to evaluate algorithm performance.

## Results

### Data Overview

This study relied on the analysis of two types of graphs – discursive graphs (noun-phrase interactions), and resonance graphs (user discursive similarity). Examples of each are shown in Figure 2.

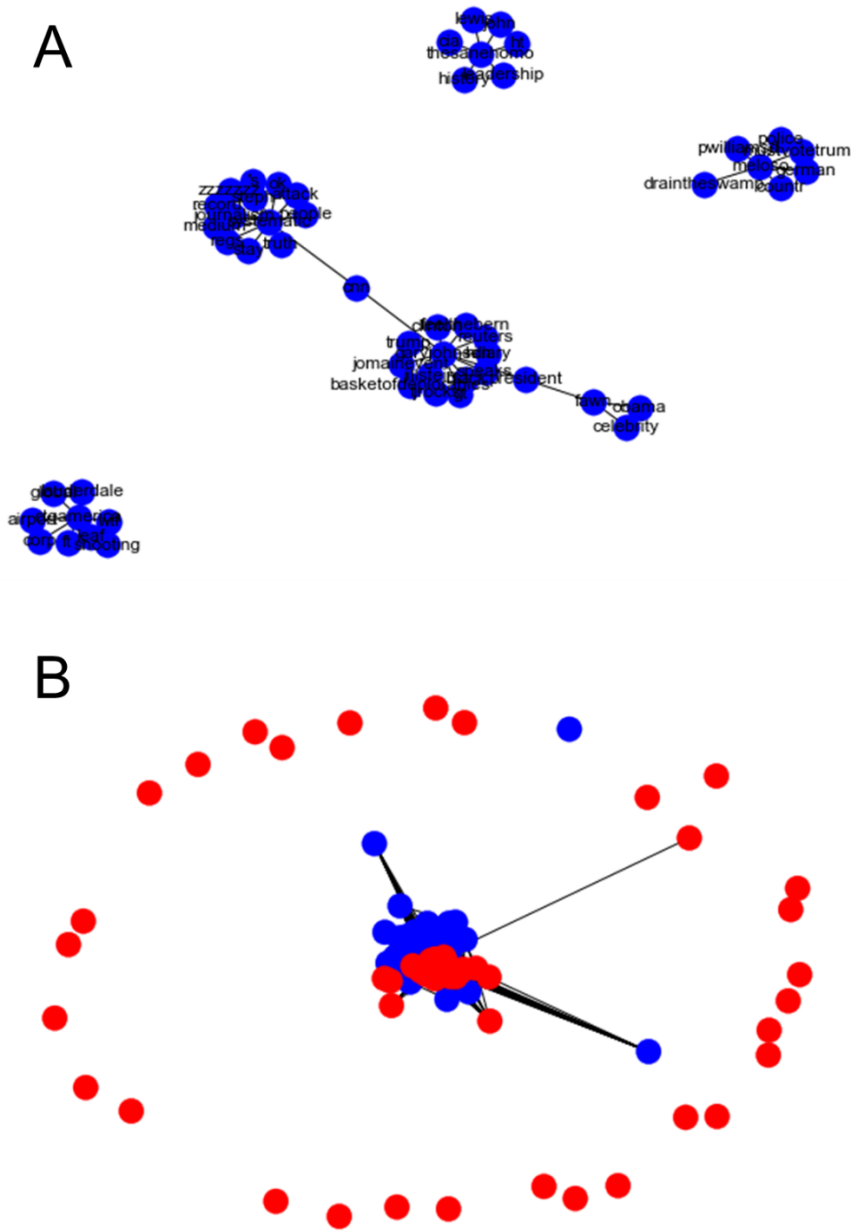

Figure 2: Representative subgraphs of a discursive graph (A) and resonance graph (B). Graph A is a discursive subgraph (induced from 50 random edges) from a known bot, where vertices represent individual words. In Graph B, blue vertices are known bots, and red vertices are control users. Graph B was induced from a slice of vertices representing equal numbers of control users and known bots. There is noticeable tight clustering of bot vertices, even in this relatively small subgraph.

**Resonance**

Because of Russian bot limitations on topics of discussion compared to control Twitter users, it is expected that Russian bots will have a high degree of discursive similarity (resonance) with each other, and a limited degree of discursive similarity with control users. An overview of the calculated resonances by interaction type is shown in Figure 3. Additionally, given the myriad topics discussed on Twitter, it is expected that a random group of Twitter users will not exhibit a high degree of intra-group discursive similarity, thus it is reasonable to assume that the control set of users represents a random body of tweets. Analysis of variance determined a p-value that is < 0.05, thus there is a statistically significant difference between the interactive sets.

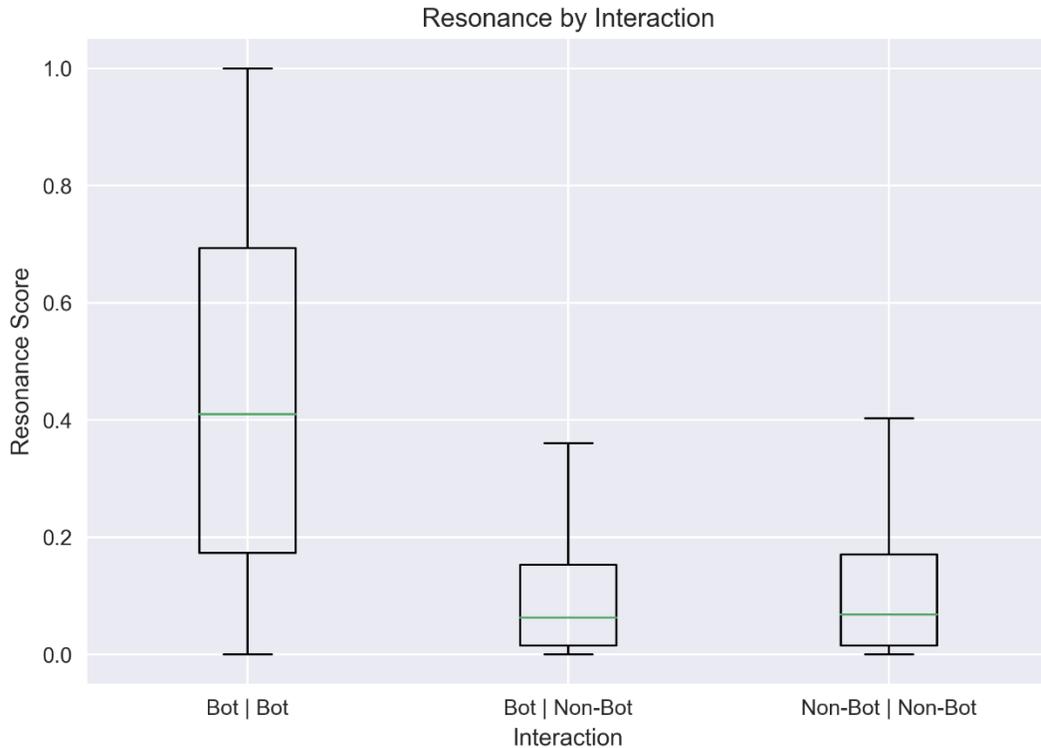

Figure 3: Resonance box plots by interaction type. Analysis of variance indicated a statistically significant difference between bot-bot interactions and other interaction types, which supports the hypothesis that bots will have greater discursive similarity with each other than with control users.

The resonance matrix for all users is shown in Figure 4. Yellow areas represent low resonance, and blue areas represent high resonance values. The black bars in the x and y axes represent known bots. As expected, the lower-right quadrant (bot-bot interactions) are highly resonant, while control-control interactions (upper-left quadrant), and control-bot interactions (upper-right and lower-left quadrants) were effectively not resonant, confirming that discursive similarity largely exists only within bot-bot interactions. This is supported by the ANOVA results.

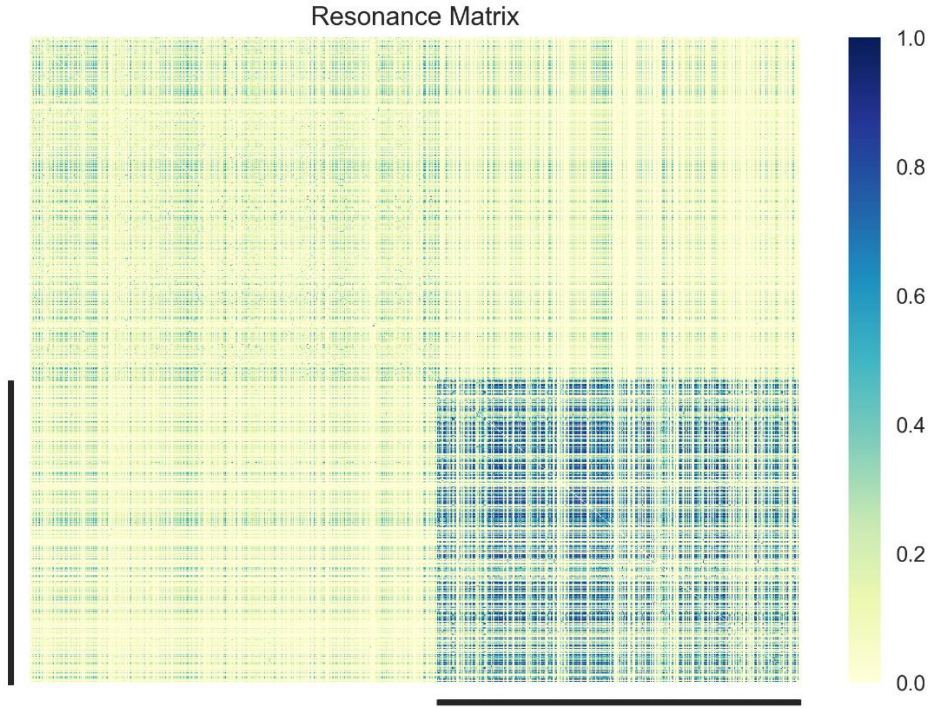

Figure 4: Colorized resonance matrix. The blue regions represent higher resonance interactions. The black bars indicate known bots. As expected, bots show a higher degree of resonance with each other than with control users. Note: This matrix is symmetric across the diagonal.

**Threshold Parameter**

Increasing the threshold parameter decreased the number of vertices considered in the clustering algorithm, as the number of edges was reduced, and therefore the number of isolated vertices tended to increase. Because the clustering algorithm only considered clusters with size > 1, more vertices would be distributed within isolated community sets, and thus not considered. This is demonstrated in Figure 5.

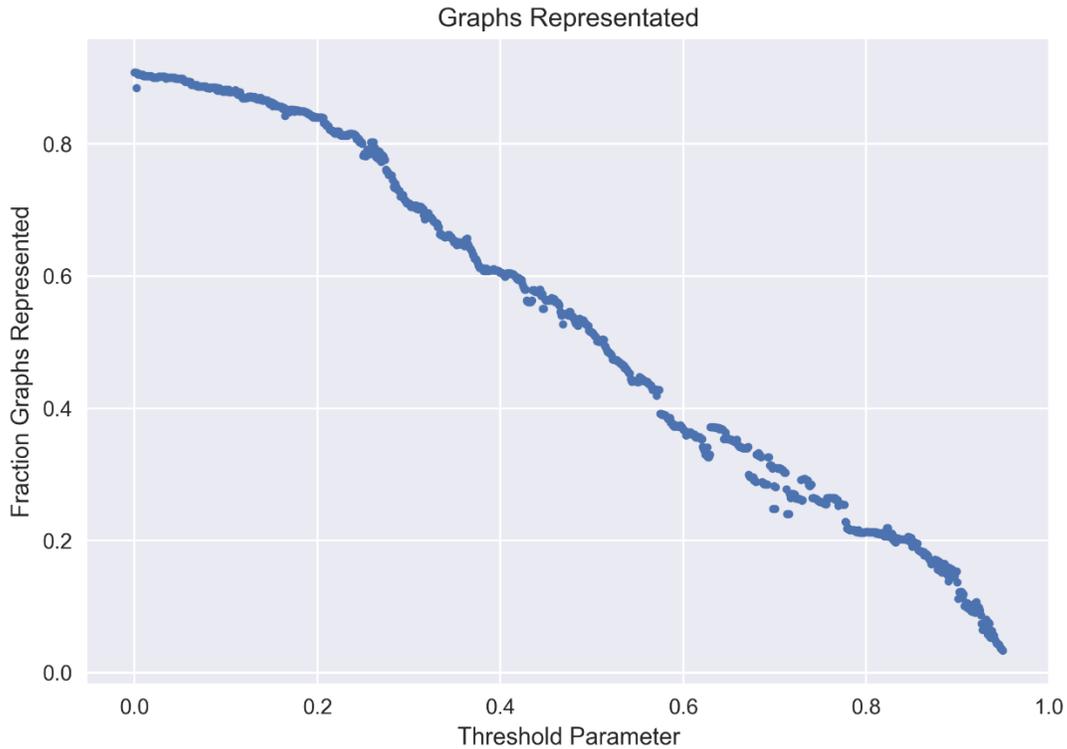

Figure 5: Represented fraction of graphs (users) by threshold parameter. As more edges are removed, the incidence of isolated vertices increases, thus more vertices are excluded from the communities.

**Matthew's Correlation Coefficient**

Matthew's Correlation Coefficient has been shown to be an effective measure of the quality of machine learning binary classification systems [12]. The formula is more robust than standard measures of accuracy and precision because it considers both true and false positives in evaluating model quality. It returns a value between -1 and 1, with 1 indicating perfect prediction, 0 indicating no predictive value, and scores approaching -1 indicating anti-predictive behavior. The formula is given in Equation 6:

$$MCC = \frac{TP \times TN - FP \times FN}{\sqrt{(TP+FP)(TP+FN)(TN+FP)(TN+FN)}} \qquad (6)$$

In this equation, *TP* is true positives, *TN* is true negatives, *FP* is false positives, and *FN* is false negatives. For every threshold parameter iteration, the Matthews Correlation Coefficient is shown in Figure 6.

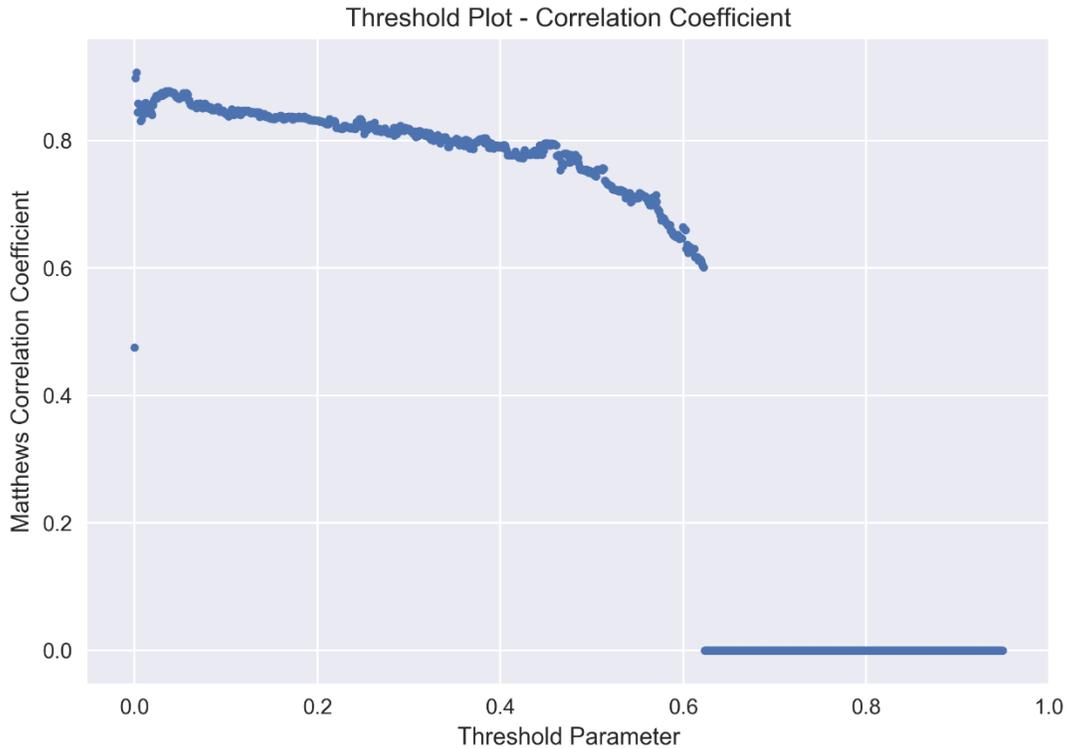

Figure 6: Matthew's Correlation Coefficient for all values of the threshold parameter ($\tau$). The maximum occurs at $\tau = 0.0019$, MCC = 0.9070, thus this is the optimal threshold parameter value for maximal predictive efficacy.

**Optimal Conditions**

Using the maximal value of Matthew's Correlation Coefficient to compute the optimal $\tau$, the confusion matrix for this data set is shown in Table 1.

|  | True Condition | |
| --- | --- | --- |
| Predicted Condition | Bot | Control |
| Bot | 0.840 | 0.064 |
| Control | 0.033 | 0.064 |

Table 1: Confusion matrix for optimal $\tau$, presented as a conditional probability table (note: numbers do not add to 1.000 due to rounding).

The sensitivity of this method at the optimal condition, i.e., the fraction of true positives over condition positives, is 95%. The probability of detecting a true positive is 84%, the probability of detecting a false negative is only 3.3%, and the probability of detecting a false positive is only 6.4%. Thus, this method is very effective at detecting bots with high discursive similarity. This method is not able to effectively detect control users (true negatives, i.e $P(predicted_{neg}|True_{neg})$), likely because of the discursive dissimilarity between control users, which limits their capacity to form appreciable clusters. However, the inability to definitively detect non-bots does not detract from the utility of this method to detect true positives (bots).

**Metric Anomalies**

It is interesting to note that there is a significant drop in the Matthew's Correlation Coefficient that occurs around $\tau = 0.62$. It is likely that this is the point where the threshold parameter is large enough where the only significant connected subgraph that exists within the larger set is the bot community, thus no true negatives (control users) are present. Given the fact that the discursive similarity is significantly greater in bot-bot interactions than in bot-control and control-control interactions, it stands to reason that increasing the threshold parameter beyond a certain point will overfit the model in favor of bot-bot interactions by excluding all or most control users.

## Discussion

With the connectedness of social media, and the speed by which disinformation can propagate through vast online social networks, it is imperative that democratic governments develop more advanced means of combating disinformation that could influence democratic processes. Detecting adversarial social network bot activity is the first step in this process. This study demonstrates that discursive similarity can be employed as an effective detection parameter for bots that propagate disinformation. It is assumed that bot topics of discussion are much narrower than topics discussed by general Twitter users, leading to greater discursive similarity between bots relative to the general Twitter conversation space. This study demonstrates that this phenomenon can be exploited for bot detection.

It would be possible to integrate this detection method into a recursive learning algorithm for bot detection. By iteratively including new users of unknown status, one can use clustering to assign the appropriate label (bot or not bot) to each interrogated user. If a bot is detected, the user can be included in all future iterations of bot detection. Thus, as bot discourse drifts over time, the model can be continuously retrained.

Due to limitations on the availability of Twitter data for research purposes, the control data set used in this study included tweets from 2010. In contrast, the bot tweets were from the 2016 election timeframe. It stands to reason that topics of discourse will drift in time, thus there is likely a greater degree of separation between the control set and the bot data set that what would exist if the sets were all sampled from the same timeframe. However, given the minimal discursive similarity between users in the control set, which will likely not change in time (unless a significant world event occurs that everyone happens to be talking about), this temporal effect is very likely minimal. While there may be more control users talking about the election and election-tangential topics (much like the disinformation bots) as we move closer to an election, this may simply manifest as a higher rate of false-positive detection (i.e., lower sensitivity) as this time gap closes. This phenomenon should be examined more closely in a subsequent study.

This leads to a second point of discussion: the incidental detection of non-bot users who frequently share posts from Russian bots (i.e, false positives). While this method is more likely to detect "bot-adjacent" users over the general Twitter population, it is likely that these users will post things about their lives other than Russian disinformation. Thus, they will exist somewhere between the Russian bots and the general twitter population in terms of discursive content, and may even form a separate community that could be resolved. While this was not specifically addressed in this study, with further development, it may be possible to use this approach to differentiate users with that level of granularity.

It is worth noting that, leading up to and during the 2016 Presidential Election, Russian bots generally propagated either far-right or far-left disinformation. Thus, one might expect to see the formation of two distinct communities of bots. It is likely that the discursive dissimilarities between left-wing and right-wing bots was too small compared to the discursive dissimilarity between bot-control interactions and all other interactions, thus this distinction was not resolved in this analysis. Future work in this area could focus on assigning content polarities ("left" or "right) to the set of bots and examine community aggregation based on that feature.

Overall, this study demonstrates an effective means of identifying discursively similar bots on Twitter, with implications for limiting the spread of disinformation that could impact democratic processes. Future work should consider expanding this concept, including the use of edge-weighted community detection algorithms to reduce or eliminate the need for association matrix thresholding. Additionally, it is worth exploring the possibility of a recursive algorithm for continuous bot detection utilizing this method.